\documentclass[preprint,12pt]{aastex}           
\usepackage{natbib,graphics}

\shorttitle{Diskoseismology Confronts Spin}
\shortauthors{Wagoner}

\begin{document}
\title{Diskoseismology and QPOs Confront Black Hole Spin}

\author{Robert V. Wagoner\altaffilmark{1}} 
\affil{Department of Physics and KIPAC, 
Stanford University, Stanford, CA 94305--4060}
 
\altaffiltext{1}{wagoner@stanford.edu} 

\begin{abstract}
We compare the determinations of the angular momentum of stellar mass black holes via the continuum and line methods with those from diskoseismology. The assumption being tested is that one of the QPOs (quasi--periodic oscillations) in each binary X--ray source is produced by the fundamental g--mode. This should be the most robust and visible normal mode of oscillation of the accretion disk, and therefore its absence should rule out diskoseismology as the origin of QPOs. The comparisons are consistent with the second highest frequency QPO being produced by this g--mode, but are not consistent with models in which one 
QPO frequency is that of the innermost stable circular orbit.
\end{abstract}


\section{Introduction}

In principle, all adiabatic perturbations of equilibrium models of accretion disks can be analyzed in terms of global normal modes. The pioneering studies of Shoji Kato and his group and the more recent work of our group have focused on accretion disks around black holes, so that no complications from boundary layers are involved. In addition, the only QPOs with fairly stable frequencies are the high frequency ones (HFQPOs) in such sources. This stability implies a fixed `cavity', which suggests gravitational trapping of the modes. 

For reviews of `relativistic diskoseismology', see \citet{k01} and \citet{w08}. A short summary of some observationally relevant results from our analyses of the low-lying spectrum, which consists of g--modes \citep{per}, c--modes \citep{swo}, fundamental p--modes \citep{osw}, and other p--modes \citep{rd4} is given by \citet{wso}. Local analyses (restricted radial interval) have also played an important role in our understanding of these perturbations \citep{kfm}. 

In section 2 we summarize the foundations and results of our approach. The predictions of black hole spin (angular momentum) are compared with those of the `continuum' and `line' methods in section 3. Section 4 contains a discussion of the viability of our model and conclusions. 

\section{Assumptions and Predictions}

We employ a fully relativistic analysis of linear hydrodynamic perturbations of geometrically thin, optically thick accretion disks. The gravitational field of the disk and its perturbations are negligible compared to that of the (Kerr metric) black hole. To simplify the analysis, we consider barotropic disks [$p=p(\rho)$], so the buoyancy frequency vanishes. This should be a good approximation since the disk is turbulent, thus close to being vertically isentropic. The equilibrium disk is taken to be described by the standard relativistic thin disk model \citep{nt,pt} [see also Section 7 of \citet{rd4}]. We have also included viscosity in the perturbation analysis \citep{ort}. The dimensionless black hole angular momentum parameter $a=cJ/GM^2$ is less than unity in absolute value, but should be positive for these black holes accreting material from a stellar companion. 

We have applied the general relativistic formalism that \citet{il} developed for perturbations of purely rotating perfect fluids. They show that one can express the Eulerian perturbations of all physical quantities through a single function proportional to the pressure perturbation $\delta p = F(r,z)\exp[i(m\phi + \sigma t)]$, where $\sigma$ is the eigenfrequency and $m$ is the axial mode number. In what follows, $\sigma<0$ and $m\ge 0$, but the same results apply with both signs reversed.
The corotation frequency $\omega = \sigma+m\Omega$, where $\Omega(r)$ is the angular velocity of the equilibrium disk.
We have adopted a weak radial WKB approximation (characteristic radial wavelength $\lambda_r \ll r$), except near the corotation resonance [which lies outside the region where the relevant modes are trapped \citep{sw}] . This has turned out to be a good approximation for essentially all of the modes.

The most robust and visible mode should be the fundamental ($m=0$, and small radial and vertical mode numbers) g--mode, for the following  reasons. It is trapped ($\omega^2 < \kappa^2$) by relativistic gravity slightly below the maximum of the radial epicyclic frequency $\kappa(r)$, which is in the hottest region of the accretion disk. Its radial extent is greater than that of the p--modes, as well as that of the c--mode for $a\gtrsim 0.1$ \citep{w08}. The inner boundary of the p-- and c--modes is near the innermost stable circular orbit (ISCO, where $\kappa=0$), which implies significant leakage of these modes into the (weakly radiating) inspiral region as well as uncertainty of the inner boundary conditions. In addition, any nonaxisymmetric mode ($m\neq 0$) can only be made visible by Doppler boosting or eclipses by the black hole.

The frequency of the fundamental g--mode lies below the maximum value of $\kappa$ by a fractional amount of approximately $h/r \sim c_s/(r\Omega)$, where $h(r)$ is the thickness of the disk and $c_s(r)$ is the speed of sound on the midplane. This splitting is proportional to the luminosity $L$ since the modulated photons come from the radiation pressure dominated region of the disk. This is valid for $0.01\lesssim L/L_{Edd}\lesssim 0.5$, in which case we obtain for the frequency $f=|\sigma|/2\pi$ the key formula [plotted in Fig.~1 of \citet{wso}]
\begin{equation}
Mf/10M_\odot = F(a)[1-\epsilon(L,a)] \; ; \quad F=71.4-246 \;\mbox{Hz}\; (a=0-1)\; , \quad \epsilon\approx 0.1L/L_{Edd}\; , \label{gmode}
\end{equation}
where $M$ is the mass of the black hole. The weak dependence of $\epsilon$ on $a$ is given by \citet{wso}; the value above is sufficiently accurate for our purposes. Initial use of this relation to estimate the spin of a few black holes was made by \citet{w08} and \citet{sw}.

For the lowest frequency nonaxisymmetric ($m=1$) g--mode, the frequency is significantly higher \citep{per}:
\begin{equation}
f(m=1)/f(m=0) = g(a) \; ; \quad g\approx 3.5-5.8 \; (a=0-1)\; . \label{mgmode}
\end{equation}
We also note that with increasing values of $m$, $f(m)\rightarrow mf(ISCO)$, the mode location $r_m\rightarrow r(ISCO)$, and its extent $\Delta r\rightarrow 0$ \citep{per}. Thus mode leakage becomes important and the observed modulation decreases, so they should be much less observable.

The orbital frequency at the ISCO is given by  
\begin{equation}
Mf(ISCO)/10M_\odot = H(a) \; ; \quad H=220-1610 \;\mbox{Hz}\; (a=0-1)\; . \label{ISCO}
\end{equation}
This frequency plays a central role in some other models of HFQPOs.

The most relevant global numerical simulations of black hole accretion disks have been carried out by \citet{rm09} within ideal hydrodynamics as well as magnetohydrodynamics (MHD), and by \citet{orm} within viscous hydrodynamics. The nonrotating black hole was represented by a modified Newtonian gravitational potential. The evolution was typically followed for about $10^2$ orbital periods of the inner disk. From power spectra at many radii, the $m=0$ g--mode was seen in the hydro simulations at the predicted frequency and radial location. It was not seen in the MHD simulations. However, because of the induced MRI turbulence, it would not be expected to be seen if it was at the same amplitude as in the hydro simulations. Because of the limited range of $\phi$ (with periodic boundary conditions) and frequency in the 3D simulations, the higher $m$ modes could not have been seen. 

In the presence of viscosity which acts hydrodynamically (via the $\alpha$ model) and perturbatively, \citet{ort} found that for most (including these g--) modes, a viscous instability is induced, with a mode growth rate of approximately $\alpha\Omega$. For $\alpha\sim 0.1$, it is not clear that there were enough orbits in the MHD simulations to allow sufficient growth. An effective turbulent viscosity (generated by the magneto-rotational instability) should be present in these (thin) accretion disks, but it is not yet known in what ways it acts like a hydrodynamic viscosity. 

We should note that \citet{abt} found no g--modes in their shearing--box MHD simulations of a limited radial region of an accretion disk. 
We also note that vertically integrated simulations [e.~g., \citet{ch}] cannot capture the g--modes.
From their MHD simulations, \citet{tv} claim that the $m=2,3,\ldots$ g--modes can grow to dominance over those of $m=0,1$ via the Rossby wave instability. However, this requires a large concentration of magnetic field between the black hole and the accretion disk. In addition, their simulation neglected the vertical structure of the disk.  There were some indications of the generation of p--modes within the MRI--induced turbulence, however.

\section{Comparison to Determinations of Black Hole Spin from the Continuum and Line Methods} 

The major methods of measuring the spin of black holes have been by (a) fitting the continuum spectrum and (b) obtaining the shape of the gravitationally redshifted wing of an iron line. An example of the use of both methods [which essentially determine the value of $r_{ISCO}(a)/M$] is provided by \citet{srm}. The continuum method requires knowledge of the black hole mass, the inclination angle, and the distance to the source; while the line method requires a knowledge of the radial dependence of the emissivity of the reflection--fluorescence line. Of course, both methods depend upon numerical simulations which capture all the relevant physical conditions in the accretion disk and `corona'.
 
\begin{deluxetable}{lcrclr}
\tablewidth{0pt}
\tablecaption{Determinations of Black Hole Spin} 
\tablehead{\colhead{Source} & \colhead{$M/M_\odot$} & \colhead{$f$(Hz)} & \colhead{$Mf/10M_\odot$} & \colhead{Spin (g--mode)} & \colhead{Spin (method)}} 
\startdata

GRS1915+105   &  14.0$\pm$4.4  &  41       &  $57\pm18$     & $0.00<a<0.27$  &  $0.54<a<0.58$\ (L)    \\
              &                &  67       &  $94\pm29$     & $0.00<a<0.72$  &  $0.97<a<0.99$\ (L)    \\
              &                & {\bf 113} & $158\pm50$     & $0.51<a<0.98$  &  $0.98<a<1.00$\ (C)    \\
              &                & {\bf 168} & $235\pm74$     & $0.82<a<1.00$  &                        \\[10pt]
XTEJ1550--564 &  $9.1\pm0.6$   &  92?      &  $84\pm6 $     & $0.13<a<0.46$  & $-0.11<a<0.71$\ (C)    \\ 
              &                & {\bf 184} & $167\pm11$     & $0.80<a<0.93$  &  $0.33<a<0.70$\ (L)    \\       
              &                & {\bf 276} & $251\pm17$     & $0.98<a<1.00$  &  $0.75<a<0.77$\ (L)    \\[10pt] 
GROJ1655--40  &  $6.3\pm0.5$   & {\bf 300} & $189\pm15$     & $0.86<a<0.98$  &  $0.65<a<0.75$\ (C)    \\     
              &                & {\bf 450} & $284\pm23$     &  No solution   &  $0.90<a<1.00$\ (L)    \\
              &                & \nodata   & \nodata        & \nodata        &  $0.97<a<0.99$\ (L)    \\[10pt]
Cyg X--1      &  $14.8\pm1.0$  &  135      & $200\pm14$     & $0.90<a<0.99$  &  $0.04<a<0.06$\ (L)    \\
              &                & \nodata   & \nodata        & \nodata        &  $0.92<a<1.00$\ (C)    \\
              &                & \nodata   & \nodata        & \nodata        &  $0.95<a<0.98$\ (L)    \\[10pt]
XTEJ1650--500 &     $5\pm2$    &  250      & $125\pm50$     & $0.07<a<0.92$  &  $0.78<a<0.80$\ (L)    \\[10pt]
XTEJ1859+226  &     $>5.4$     &  190      & $>103$         & $0.46<a$       &                        \\[10pt]
H1743--322    &    \nodata     & {\bf 165} & \nodata        & \nodata        &  $-0.3<a<0.7$\ (C)     \\
              &                & {\bf 241} & \nodata        & \nodata        &                        \\

\enddata
\tablecomments{The black hole masses are from (in order) \citet{hg,osm,gbo,oma,omr,ccs}. The QPO frequencies are from the compilations of \citet{rm06,rm11}, with the 3/2 ratios in boldface. The 92 Hz QPO is controversial. The determinations of spin from the continuum (C) method are from (in order) \citet{msn,srm,smn,gmr,smr}, while those from the line (L) method are from \citet{bmf}(first two);\ \citet{srm,mrf,rfr};\ \citet{mrf}(next two);\ \citet{fwm,mrf}.}
\end{deluxetable}

In Table 1 we compare determinations of black hole spin via these two methods with that from the fundamental g--mode [equation (\ref{gmode})], for all sources with high--frequency QPOs and at least one determination of spin. We have not included the errors in the QPO frequencies, since they are typically smaller than the errors in $M$. Note that for all sources with at least two HFQPOs, the two highest frequencies have a ratio very close to 3/2, which may provide an important clue to their origin \citep{ak}. A key source is GROJ1655--40, since the highest frequency (requiring $a>1$) cannot be due to the g--mode. However, the g--mode determination of spin from the lower frequency QPO is consistent with those from the line method, but somewhat above that from the continuum method. Note, however, that these two methods do not give compatible results. 

We therefore advance the hypothesis that the lower frequency member of the four 3/2 pairs is produced by the g--mode, and tentatively  assume that is also true for the three single HFQPOs. Then we see that for GRS1915+105, this assumption predicts a spin which is consistent with all three other determinations. For XTEJ1550--564, the g--mode spin determination is slightly higher than those from the other methods. For Cygnus X--1, our determination of $a$ is in agreement with that from the continuum method and a recent result from the line method [which also agrees with the preferred value from the line method of \citet{ddw}]. The earlier result from the line method is in strong disagreement with the other results. For XTEJ1650--500, the g--mode result is consistent with that from the only other spin determination, although the value of the black hole mass is uncertain. Unfortunately, no mass values are available for the other source with a 3/2 pair of HFQPOs, H1743--322. The range of spin indicated corresponds to using a probability distribution of the mass, indicating $5\lesssim M/M_\odot \lesssim 15$.

For the $m=1$ nonaxisymmetric g--mode, we see from equations (\ref{gmode}) and (\ref{mgmode}) that $Mf/10M_\odot \approx 250-1430$ Hz for $a=0-1$. We note [from equation (\ref{ISCO})] that the ISCO frequency has a similar range of values. Thus from the observed values in Table 1 we see that only very small values of $a$ are allowed, and only for the highest frequency QPO. Thus we can rule out these two models from the comparison with the relatively large values of spin required by most of the continuum and line determinations. 

\section{Discussion and Conclusions}

An important first step in including the effects of magnetic fields within diskoseismology was taken by \citet{fl}. They found (within a local WKB analysis) that a vertical component of magnetic field could affect the trapping of the $m=0$ g--mode when its pressure was at least $\sim 10^{-3}$ times the gas pressure. The inner trapping radius of the mode became that of the ISCO, and the mode frequency became larger than the maximum value of $\kappa(r)$. One is then led to a scenario in which the stable QPO only exists when the magnetic field is small enough. However, this is consistent with the observation that the HFQPOs appear only sporadically \citep{mr,rm06}.

A key question is whether the accretion disk can be analyzed in terms of perturbations (normal modes plus turbulence) to a stationary structure. The turbulence could provide the driving force and (anti) damping of the modes, within a weakly nonlinear analysis. One is naturally tempted to contemplate the outcome of the mode growth. Will enough nonlinearity be induced to lead to significant mode--mode coupling and related effects? Some evidence that nonlinear effects should be weak is the observed lack of dependence of the HFQPO frequencies on their amplitude. 

What perturbations (or nonlinear structures) correspond to the QPOs not associated with the fundamental g--mode? One would like to extend analyses such as the test particle resonance model of \citet{ak} or the wave coupling model of \citet{k08} to this problem. This has been done by \citet{ho}, who analyzed weak nonlinear coupling between epicyclic modes \citep{bsa} in slender tori, extending the earlier work of \citet{akk} to global normal modes. When we apply a weak coupling analysis to our thin disk modes, a problem arises. There is a selection rule, $\sum m=0$ over coupled modes. To lowest order, in which three mode couplings are considered, the only simple way to obtain the 3/2 ratio between the two highest frequencies is by use of the $m\neq0$ g--modes. However, as shown above, such modes are ruled out observationally. We are exploring other possibilities. We note that \citet{rym} obtained a 3/2 ratio between p--modes in small accretion tori. However, the frequencies depend upon the size of the torus and the speed of sound.  

We anticipate that numerical simulations will give us more insight into the physical conditions within the accretion disk and `corona'. However, the fact that no simulations have produced the observed spectra of QPOs indicates that they are incomplete.
Hopefully the ASTROSAT satellite (the successor to RXTE, whose archives will still be providing important data for analysis) will soon provide new black hole sources. In the future, the increased area of the proposed LOFT satellite should also help us solve some of the mysteries of the QPOs. More precise determinations of black hole masses and spins, as well as more robust determinations of QPO frequencies, will hopefully provide a more definitive test of diskoseismology. 

\acknowledgments

We thank Alex Silbergleit and Manuel Ortega-Rodr\'{\i}guez for their previous contributions to various aspects of this research, 
and the referee for supplying two useful references.


\begin{thebibliography}{}

\bibitem[Abramowicz \& Klu\'{z}niak(2001)]{ak} Abramowicz, M.A. \& Klu\'{z}niak, W. 2001, \aap, 374, L19
\bibitem[Abramowicz et al.(2003)]{akk} Abramowicz, M.A., Karas, V., Klu\'{z}niak, W., Lee, W.H., \& Rebusco, P. 2003, \pasj, 55, 467
\bibitem[Arras et al.(2006)]{abt} Arras, P., Blaes, O., \& Turner, N.J. 2006, \apjl, 645, L65
\bibitem[Blaes et al.(2007)]{bsa} Blaes, O.M., \v{S}r\'{a}mkov\'{a}, E., Abramowicz, M.A., Klu\'{z}niak, W., \& Torkelsson, U. 2007, \apj, 665, 642
\bibitem[Blum et al.(2009)]{bmf} Blum, J.L., Miller, J.M., Fabian, A.C., et al. 2009, \apj, 706, 60
\bibitem[Chan(2009)]{ch} Chan, C.-K. 2009, \apj, 704, 68
\bibitem[Corral-Santana et al.(2011)]{ccs} Corral-Santana, J.M., Caseras, J., Shahbaz, T., et al. 2011, \mnras, 413, L15
\bibitem[Duro et al.(2011)]{ddw} Duro, R., Dauser, T., Wilms, J., et al. 2011, \aap, 533, L3
\bibitem[Fabian et al.(2012)]{fwm} Fabian, A.C., Wilkins, D., Miller, J.M., et al. 2012, arXiv:1204.5854
\bibitem[Fu \& Lai(2009)]{fl} Fu, W. \& Lai, D. 2009, \apj, 690, 1386
\bibitem[Gou et al.(2011)]{gmr} Gou, L., McClintock, J.E., Reid, M.J., et al. 2011, \apj, 742, 85 
\bibitem[Greene et al.(2001)]{gbo} Greene, J., Bailyn, C.D., \& Orosz, J.A. 2001, \apj, 554, 1290
\bibitem[Harlaftis \& Greiner(2004)]{hg} Harlaftis, E.T. \& Greiner, J. 2004, \aap, 414, L13
\bibitem[Hor\'{a}k(2008)]{ho} Hor\'{a}k, J. 2008, \aap, 486, 1
\bibitem[Ipser \& Lindblom(1992)]{il} Ipser, J.R. \& Lindblom, L. 1992, \apj, 389, 392
\bibitem[Kato(2001)]{k01} Kato, S. 2001, \pasj, 53, 1 
\bibitem[Kato(2008)]{k08} Kato, S. 2008, \pasj, 60, 111 
\bibitem[Kato et al.(2008)]{kfm} Kato, S., Fukue, J., \& Mineshige, S. 2008, Black-Hole Accretion Disks: Towards a New Paradigm (Kyoto: Kyoto University Press)
\bibitem[McClintock \& Remillard(2006)]{mr} McClintock, J.E \& Remillard, R.A. 2006, in Compact Stellar X-ray Sources, ed. W. Lewin \& M. van der Klis (Cambridge: Cambridge Univ. Press) p. 157  
\bibitem[McClintock et al.(2006)]{msn} McClintock, J.E., Shafee, R., Narayan, R., et al. 2006, \apj, 652, 518
\bibitem[Miller et al.(2009)]{mrf} Miller, J.M., Reynolds, C.S., Fabian, A.C., Miniutti, G., \& Gallo, L.C. 2009, \apj, 697, 900
\bibitem[Novikov \& Thorne(1973)]{nt} Novikov, I.D. \& Thorne, K.S. 1973, in Black Holes, ed. C. DeWitt \& B.S. DeWitt (New York: Gordon and Breach)
\bibitem[O'Neill et al.(2009)]{orm} O'Neill, S.M., Reynolds, C.S., \& Miller, M.C. 2009, \apj, 693, 1100
\bibitem[Orosz et al.(2004)]{omr} Orosz, J.A., McClintock, J.E., Remillard, R.A., \& Corbel, S. 2004, \apj, 616, 376
\bibitem[Orosz et al.(2011a)]{oma} Orosz, J.A., McClintock, J.E., Aufdenberg, J.P., et al. 2011, \apj, 742, 84
\bibitem[Orosz et al.(2011b)]{osm} Orosz, J.A., Steiner, J.F., McClintock, J.E., et al. 2011, \apj, 730, 75
\bibitem[Ortega-Rodr\'{\i}guez et al.(2002)]{osw} Ortega-Rodr\'{\i}guez, M., Silbergleit, A.S., \& Wagoner, R.V. 2002, \apj, 567, 1043
\bibitem[Ortega-Rodr\'{\i}guez et al.(2008)]{rd4} Ortega-Rodr\'{\i}guez, M., Silbergleit, A.S., \& Wagoner, R.V. 2008, Geophys. Astrophys.~Fluid Dyn., 102, 75
\bibitem[Ortega-Rodr\'{\i}guez \& Wagoner(2000)]{ort} Ortega-Rodr\'{\i}guez, M. \& Wagoner, R.V. 2000, \apj, 537, 922
\bibitem[Page \& Thorne(1974)]{pt} Page, D.N. \& Thorne, K.S. 1974, \apj, 191, 499
\bibitem[Perez et al.(1997)]{per} Perez, C.A., Silbergleit, A.S., Wagoner, R.V., \& Lehr, D.E. 1997, \apj, 476, 589 
\bibitem[Reis et al.(2009)]{rfr} Reis, R.C., Fabian, A.C., Ross, R.R., \& Miller, J.M. 2009, \mnras, 395, 1257
\bibitem[Remillard \& McClintock(2006)]{rm06} Remillard, R.A. \& McClintock, J.E. 2006, \araa, 44, 49
\bibitem[Remillard \& McClintock(2011)]{rm11} Remillard, R.A. \& McClintock, J.E. 2011, private communications
\bibitem[Reynolds \& Miller(2009)]{rm09} Reynolds, C.S. \& Miller, M.C. 2009, \apj, 692, 869
\bibitem[Rezzolla et al.(2003)] {rym} Rezzolla, L., Yoshida, S., Maccarone, T.J., \& Zannoti, O. 2003, \mnras, 344, L47
\bibitem[Shafee et al.(2006)]{smn} Shafee, R., McClintock, J.E., Narayan, R., et al. 2006, \apjl, 636, L113
\bibitem[Silbergleit \& Wagoner(2008)]{sw} Silbergleit, A.S. \& Wagoner, R.V. 2008, \apj, 680, 1319 
\bibitem[Silbergleit et al.(2001)]{swo} Silbergleit, A.S., Wagoner, R.V., \& Ortega-Rodr\'{\i}guez, M. 2001, \apj, 548, 335 
\bibitem[Steiner et al.(2011)]{srm} Steiner, J.F., Reis, R.C., McClintock, J.E., et al. 2011, \mnras, 416, 941
\bibitem[Steiner et al.(2012)]{smr} Steiner, J.F., McClintock, J.E., \& Reid, M.J. 2012, \apjl, 745, L7
\bibitem[Tagger \& Varni\'ere(2006)]{tv} Tagger, M. \& Varni\'ere, P. 2006, \apj, 652, 1457
\bibitem[Wagoner(2008)]{w08} Wagoner, R.V. 2008, New Astron. Rev., 51, 828
\bibitem[Wagoner et al.(2001)]{wso} Wagoner, R.V., Silbergleit, A.S., \& Ortega-Rodr\'{\i}guez, M. 2001, \apjl, 559, L25

\end{thebibliography}
\end{document}